\documentclass[aps,prl,twocolumn,groupedaddress,amsmath,slashbox,amssymb]{revtex4-1}
\usepackage{graphicx}  % needed for figures
\usepackage{dcolumn}   % needed for some tables
\usepackage{bm}        % for math
\usepackage{verbatim}   % for math
\usepackage{hyperref}

\begin{document}

\title{Multiplexed readout of four qubits in 3D cQED architecture using broadband JPA}
\author{Suman Kundu$^{1}$}
\author{Nicolas Gheeraert$^{1,2}$}
\altaffiliation{Current address: Research Center for Advanced Science and Technology (RCAST), The University of Tokyo, Meguro-ku, Tokyo 153-8904, Japan}
\author{Sumeru Hazra$^{1}$}
\author{Tanay Roy$^{1}$}
\altaffiliation{Current address: Department of Physics and James Franck Institute, University of Chicago, Chicago, Illinois 60637, USA}

\author{Kishor V. Salunkhe$^{1}$, Meghan P. Patankar$^{1}$}
\author{R. Vijay$^{1}$}
\email{Corresponding author: r.vijay@tifr.res.in}
%\author{Suman Kundu$^{1}$, Nicolas Gheeraert$^{1,2}$, Sumeru Hazra$^{1}$, Tanay Roy$^{1}$, Kishor V. Salunkhe$^{1}$, Meghan P. Patankar$^{1}$, and R. Vijay$^{1}$}
\affiliation{$^{1}$
   Department of Condensed Matter Physics and Materials Science,\\
   Tata Institute of Fundamental Research,\\
   Homi Bhabha Road, Mumbai 400005, India
   }
\affiliation{$^{2}$
		Institut N\'eel, CNRS and Universit\'e Grenoble-Alpes, F-38042 Grenoble, France
}
\date{\today}

\begin{abstract}
We propose and demonstrate a frequency-multiplexed readout scheme in 3D cQED architecture. We use four transmon qubits coupled to individual rectangular cavities which are aperture-coupled to a common rectangular waveguide feedline. A coaxial to waveguide transformer at the other end of the feedline allows one to launch and collect the multiplexed signal. The reflected readout signal is amplified by an impedance engineered broadband parametric amplifier with 380 MHz of bandwidth. This provides us high fidelity single-shot readout of multiple qubits using compact microwave circuitry, an efficient way for scaling up to more qubits in 3D cQED.
\end{abstract}

\maketitle

High fidelity quantum measurements are a crucial part of a scalable quantum computing architecture \cite{chuang_book}. Apart from determining the state of a quantum register, they are also important for accurate state initialization and quantum error correction. In superconducting circuit technology, the combination of circuit-quantum electrodynamics (cQED) \cite{Blais-CPB,wallraff-cqed} with ultra-low noise Josephson Parametric Amplifiers \cite{Hatridge-JPA, lehnert-paramp} ushered in the era of fast, high fidelity measurements \cite{sup-qubit-review-science}. This led to the observation of quantum jumps \cite{quantum-jump-vijay} and quantum feedback \cite{Vijay-stabilizing-rabi} and enabled accurate state initialization \cite{Siddiqi-heralding} and error detection \cite{IBM-errordet-4qubit,riste2015detecting,Martinis_9xmon_nature}. Combined with the nearly six orders of magnitude improvement in coherence times \cite{sup-qubit-review-science}, superconducting circuits have emerged as one of the leading candidates for building scalable quantum computers. While several proof-of-principle demonstrations of algorithms exist, a practical quantum advantage has not yet been demonstrated due to the small number of qubits used in the quantum processors. Multi-qubit quantum processors face challenges like variability in device parameters across a chip, efficient use of cryogenic microwave resources and reduction of gate fidelity due to unwanted crosstalk. A considerable fraction of recent research efforts is focused to solve these scalability related hurdles.

An important step in developing a scalable quantum processor is the implementation of resource-efficient measurement of large number of qubits with  high-fidelity. Typical setups require two cryogenic lines to control (input) and measure (output) a single qubit. Further, a third cryogenic line is often needed to bias the first stage parametric amplifier in the output chain. Clearly this is not practical solution beyond a few qubits. One solution is to use frequency multiplexing for the measurement lines since the typical bandwidth of readout cavities is about 5 {\textendash} 10 MHz. This idea has been successfully used in 2D cQED architecture where several readout resonators coupled to a common feedline are probed using multi-frequency pulses \cite{Ustinov-freq-division,phasequbit,multiplexedrdtvion,Martinis-fast-accurate-state-measurement,Wallraff-multiplexed-rdt}. This approach allows one to perform selective readout of any subset of qubits by choosing corresponding frequency components in the readout pulse. Moreover, the multiple frequency components can be generated using one microwave generator modulated by a high bandwidth arbitrary waveform generator which further helps in minimizing resources per qubit. Some architectures have also included Purcell protection \cite{Reed-purcell} by incorporating appropriate filters in the common feed-line \cite{Martinis-fast-accurate-state-measurement} or by putting individual Purcell filters for each readout cavities \cite{Wallraff-multiplexed-rdt}. While experiments with up to four qubits have been performed in 3D cQED architecture \cite{RIPgate, multi-qubit-3d-yale}, they have not exploited frequency multiplexing for measurement so far. Further, the modular nature of the 3D cQED architecture can help in solving several scalability related issues to construct a high quality small-scale processor in the near term.

\begin{figure}[t]
	\centering
	\includegraphics[width=0.45\textwidth]{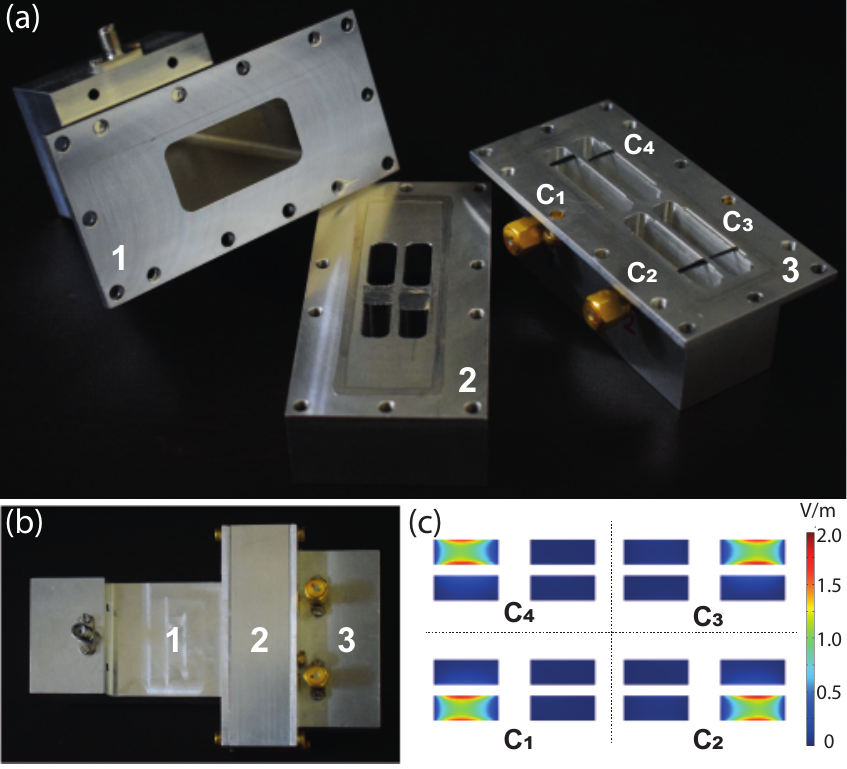} 
	\vspace{-5pt}
	\caption{(a) The four cavity structure is fabricated using Aluminum in three separate pieces. The piece marked 1 is the rectangular waveguide part with a coaxial port at one end. Piece 2 comprises one half of the readout cavities along with the coupling apertures. Piece 3 comprises of the second half of the readout cavities (C$_{1-4}$) with weakly coupled ports for separate qubit excitation if needed (not used here). (b) Fully assembled structure with each interface sealed with Indium; (c) Profile of electric field amplitude at the aperture hole between the  waveguide and readout cavities when $10^{-9}$ Watt of excitation is given at the waveguide port. The EM simulation is carried out in COMSOL\textsuperscript{\textregistered} at each cavity's resonant frequency}
	\label{fig:quadcavsimulation}
\end{figure}

In this letter, we demonstrate a multiplexed readout scheme for transmon qubits \cite{transmon_theory} in 3D cQED architecture \cite{Paik-3d-transmon} using a rectangular waveguide feed-line as both the multiplexing element \cite{waveguide_fluxonium} and a Purcell filter \cite{, 3DWaveguide}. Four rectangular readout cavities are aperture-coupled to a single rectangular waveguide as shown in Fig.~\ref{fig:quadcavsimulation}. The cavities are composed of two halves so that the qubits can be placed at the antinode of the field. A coaxial to waveguide transformer at the other end of the waveguide helps address the four separate cavities from a single line. We measured the cavities in reflection and used an impedance-engineered broadband Josephson Parametric Amplifier (JPA)\cite{BBparamp} to achieve simultaneous single-shot fidelities greater than 98\%.

\begin{figure}[t]
		\centering
		\includegraphics[width=0.5\textwidth]{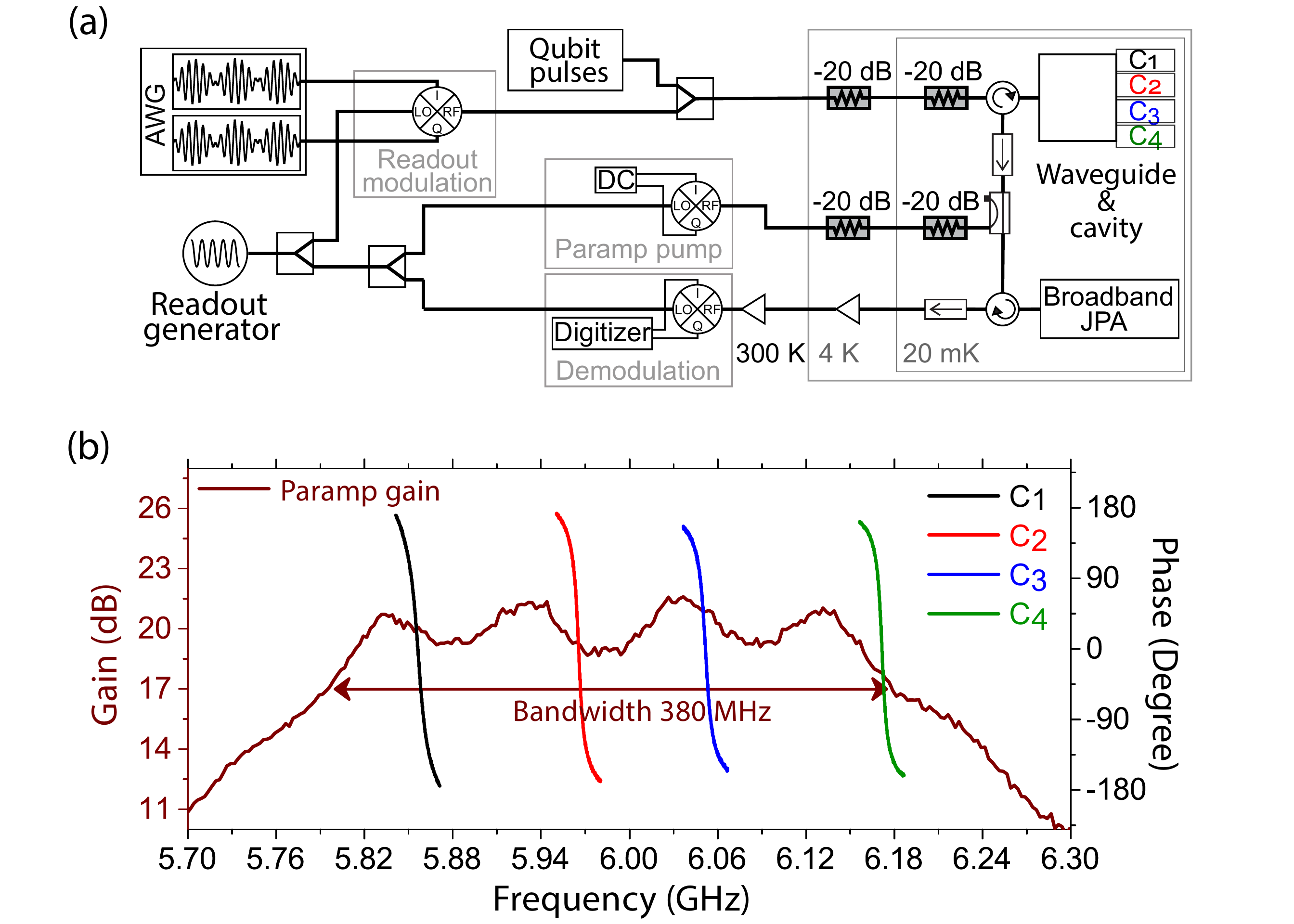} 
		\vspace{-5pt}
		\caption{(a) Setup: Readout tones are generated by sideband modulation technique from one generator which is used for paramp pump as well. The reflected signal is amplified, demodulated and then digitized; (b) Measured gain profile as a function of signal frequency is shown. 20 dB gain and 380 MHz of bandwidth is achieved at pump frequency 5.984 GHz. Reflected phase responses of four readout cavities (C$_{1-4}$) as measured from the waveguide port are plotted with respect to the right axis. }
		\label{fig:gain}
\end{figure}

The cavity frequencies are chosen with a spacing of 70 {\textendash} 100 MHz to fit within the band of the JPA as shown in Fig.~\ref{fig:gain}. The linewidths ($\kappa/2\pi$) of the readout cavities are adjusted by changing the aperture hole sizes. We first use finite element electromagnetic simulations in COMSOL\textsuperscript{\textregistered} to set the coupling aperture size. Final adjustment is done by tuning the aperture size between each cavity and the waveguide using Aluminum tape to get linewidths in the range 5 {\textendash} 10 MHz. For obtaining a clean response, it is also important to optimize the coax to waveguide transition by fine tuning the length of the coaxial center conductor inserted inside the waveguide.

To readout multiple qubits simultaneously we generate multi-frequency probe pulse(addressing all four cavities) using sideband modulation technique from a central microwave tone. This same central tone is also used as pump to bias the parametric amplifier (see Fig.~\ref{fig:gain} (a)). The reflected output signal from the waveguide port is amplified by a broadband Josephson parametric amplifier with an average gain of 20 dB in the relevant bandwidth of 380 MHz around 5.985 GHz and near quantum-limited noise. After several stages of amplification the output readout signal is down-converted with respect to the central tone and then digitized after filtering out unwanted frequency components. All cavity signals are finally demodulated in software to recover the amplitude and phase response of each individual cavity. For qubit excitations, we used the same input line as for readout tone, but one can also use separate weakly coupled ports to each individual readout cavity.

\begin{table*}[t]
	\begin{tabular}{c c c c c c c c c c c} 
		\hline
		\hline
		Cavity & Cavity & Qubit & Anharmoni- & Qubit-cavity & T$_1$ & T$_{Ramsey}$ & T$_{Echo}$ & Individual & Simultaneous  \\
		frequency & linewidth & frequency & city & coupling &  &  &  & readout  & readout \\
		${\omega_c}/{2\pi}$ (GHz) & ${\kappa}/{2\pi}$ (MHz) & ${\omega_q}/{2\pi}$ (GHz) & ${\alpha}/{2\pi}$ (MHz) & ${g}/{2\pi}$ (MHz) & ($\mu$s) & ($\mu$s) & ($\mu$s) & fidelity (\%) & fidelity (\%)\\
		\hline
		5.856 & 8.4 & 3.752 & -318 & 126 & 50.1 & 2.1 & 3.1 & 98.36 & 98.05 &\\ 
		
		5.966 & 5.2 & 4.122 & -306 & 112 & 44.1 & 1.4 & 2.7 & 98.72 & 98.57 &\\
		
		6.052 & 7.1 & 4.880 & -292 & 91 & 18.8 & 2.7 & 3.0 & 98.39 & 98.07 &\\
		
		6.172 & 5.4 & 5.278 & -297 & 81 & 25.6 & 2.6 & 2.6 & 98.74 & 98.68 &\\ [0ex]
		\hline
		\hline
	\end{tabular}
	\caption{Measured device parameters and coherence numbers and readout fidelities of four qubits (Q$_1$, Q$_2$, Q$_3$ and Q$_4$) in their corresponding readout cavities (C$_1$, C$_2$, C$_3$ and C$_4$ respectively).}
	\label{table:deviceparameter}
\end{table*}

\begin{figure}[t]
	\centering
	\includegraphics[width=0.45\textwidth]{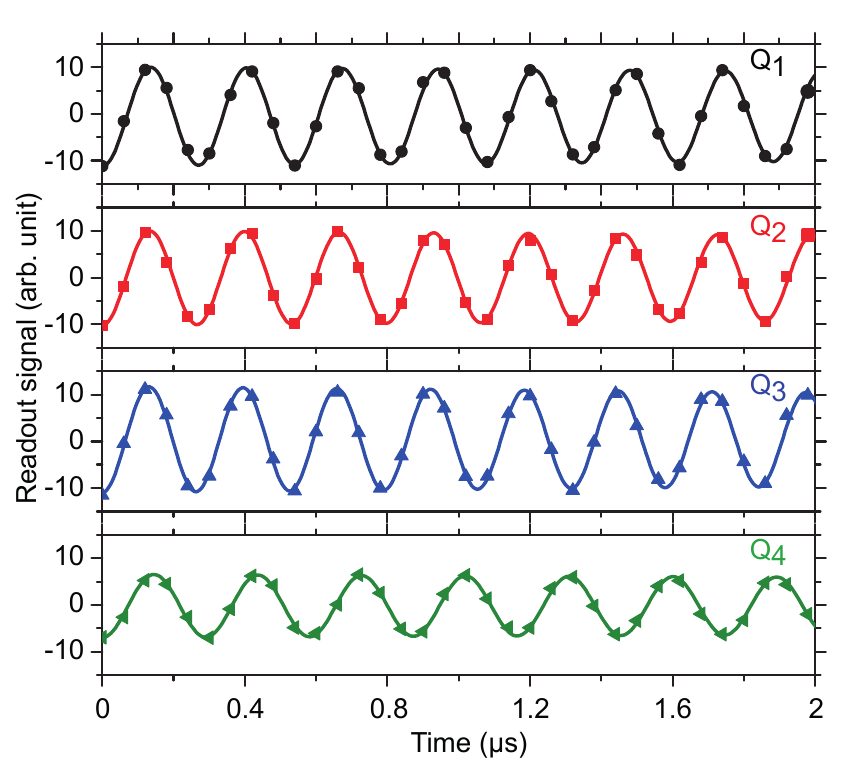} 
	\vspace{-5pt}
	\caption{Simultaneous Rabi
		oscillations of Q$_{1-4}$ as a function of the corresponding control pulse duration. The signal amplitude of Q$_4$ is smaller due to a smaller dispersive shift and reduced JPA gain at 6.172 GHz}
	\label{fig:rabi}
\end{figure}

At 30 mK temperature, all readout cavities were over-coupled and their frequencies were well-separated from each other with loaded linewidth ($\kappa/2\pi$) in the intended range of 5 {\textendash} 8 MHz. We first characterized each qubit-cavity system separately. The relevant device parameters were then experimentally determined and are listed in Table \ref{table:deviceparameter}. The values are typical for 3D transmons in cQED architecture. These were followed by time domain measurements on all qubits to determine the coherence properties (Table \ref{table:deviceparameter}). We then used the multiplexing technique to carry out simultaneous measurements of Rabi oscillations on all four qubits and the data is shown in Fig.~\ref{fig:rabi}. State-dependent dispersive shifts of all four qubits were determined using Ramsey fringe experiments in presence of different numbers of cavity photons \cite{Vijay-stabilizing-rabi}. Extracted dispersive shifts were in the range 1.19 {\textendash} 1.76 MHz, in good agreement with the dispersive shifts calculated from coupling, detuning and anharmonicity.

\begin{figure}[t]
	\centering
	\includegraphics[width=0.45\textwidth]{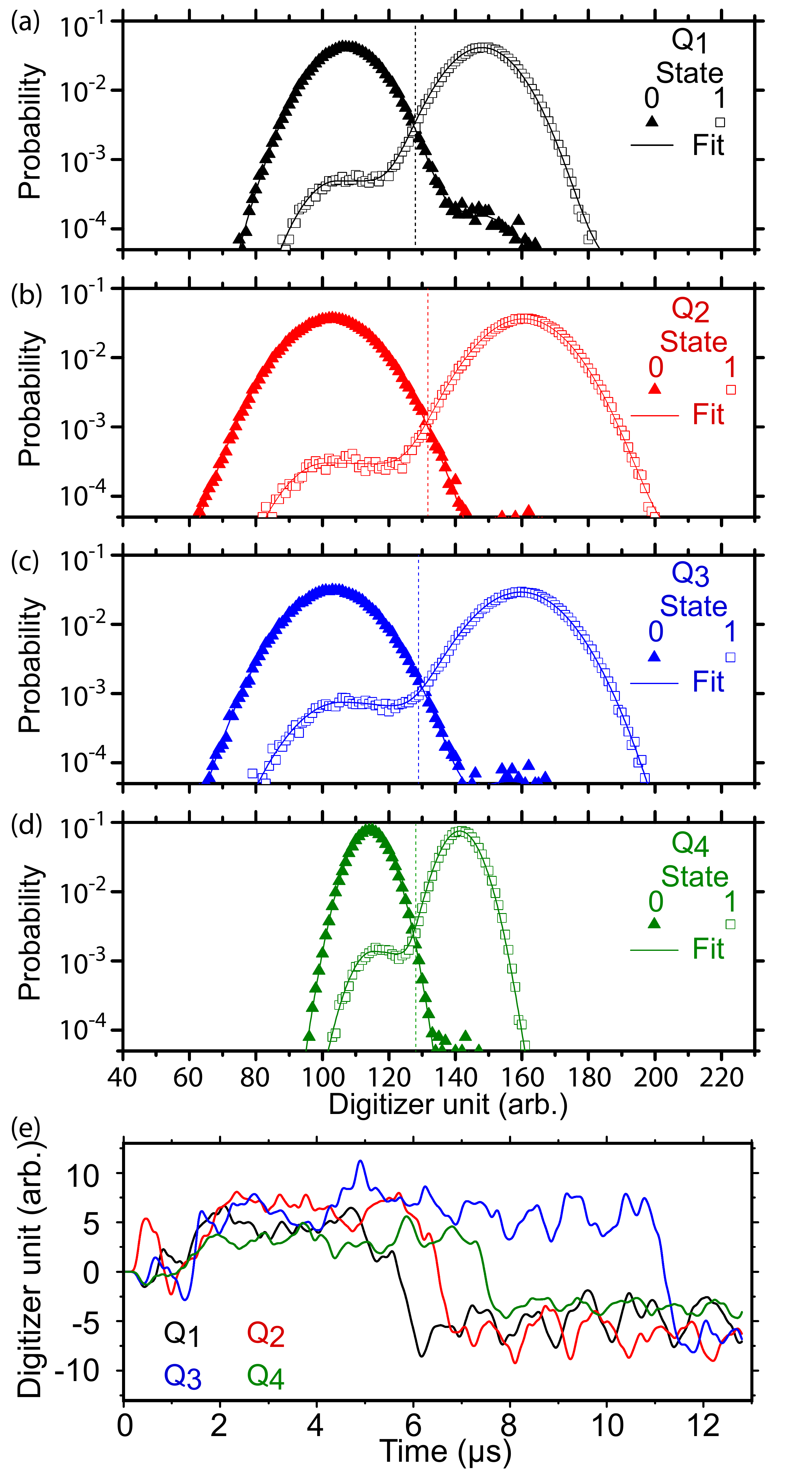} 
	\vspace{-5pt}
	\caption{Measurement histograms: (a), (b), (c) and (d) are the histograms of integrated single-shot readout signal of Q$_1$, Q$_2$, Q$_3$ and Q$_4$ respectively prepared with a $\pi$-pulse (empty square) and without (solid triangle) obtained from $3\times10^5$ measurements.The histograms are fit to a weighted sum of two Gaussian. The vertical dashed lines indicate the threshold used to discriminate state 0 and 1. (e) Quantum jump: After the qubits are initialized to their excited states, continuous strong measurement tones are applied and the reflected readout signal is digitized. Abrupt quantum jumps from the excited state to the ground state are clearly visible with no evidence of measurement crosstalk.}
	\label{fig:hist}
\end{figure}

We then carried out single-shot measurements to determine readout fidelity for all four qubits.  Each experimental sequence starts with a measurement pulse to herald the qubit in the ground state \cite{Siddiqi-heralding} and discard the data points which collapse the qubit to the excited state ($\sim$ 3\% {\textendash} 5\%). With the qubit prepared in the ground (no pulse on the qubit) or excited state ($\pi\ pulse$ on the qubit), we excite the cavity with the readout tone and integrate the reflected signal. This is repeated 3 x $10^5$ times to form histograms of the cavity response for the ground and excited state respectively. We optimize the measurement time, power and phase of the readout tone such that the overlap between the ground and excited state histograms is minimal to ensure maximum readout fidelity.  Simultaneously measured histograms for the ground and excited states for all four qubits gave high measurement fidelities ($\sim 98.5\%$) and are shown in Fig. \ref{fig:hist} (a)-(d). As shown in Table~\ref{table:deviceparameter}, these fidelities are comparable to those obtained from measuring each qubit-cavity system separately without multiplexing.
 
High fidelity single-shot measurement was further confirmed by observing quantum jumps in simultaneously taken readout traces. After preparing all qubits in their excited states, we turn on the measurement pulses for four cavities simultaneously and acquire the cavity response. As shown in Fig.~\ref{fig:hist} the traces corresponding to four qubits show clear jumps from excited to ground state and with no evidence of crosstalk.

We studied crosstalk effects on one qubit arising from both qubit and readout drives applied on other qubits. This was done by performing Ramsey fringe experiment in presence of the crosstalk drive. The crosstalk due to other qubit drives was negligible and could be further suppressed by using the weakly coupled ports in each cavity. We only found significant crosstalk between cavity 3 and 4 which led to qubit frequency and decay constant of qubit 3(4) to shift by 0.3(0.4) MHz and 4(5) $\mu$s respectively due to readout drive on cavity 4(3). This crosstalk can be reduced by minor modifications to the cavity design.

In conclusion, we have demonstrated frequency-multiplexed readout scheme in 3D cQED architecture using an impedance-engineered broadband parametric amplifier. We simultaneously measure four transmon qubits in four individual cavities coupled to a single rectangular waveguide.  A common input line for qubit and cavity excitations and  more importantly a single amplifier chain is used resource efficiently without sacrificing measurement bandwidth or qubit coherence with negligible crosstalk. Further increase of the number of qubits per measurement line would require careful optimization of cavity design and improvement of JPA dynamic range.

\textit{Acknowledgments:-} This work is supported by the Department of Atomic Energy of the Government of India. R.V. acknowledges support from the Department of Science and Technology, India via the Ramanujan Fellowship and Nano Mission. N.G. acknowledges support from the Fondation Nanosciences de Grenoble and Raman Charpak Fellowship. We acknowledge the TIFR nanofabrication facility.

%\bibliography{multiplexedreadout}

%

\end{document}